\documentclass[useAMS,usenatbib]{mnras}

\usepackage{newtxtext,newtxmath}
\usepackage[T1]{fontenc}
\usepackage{ae,aecompl}

\usepackage{graphicx}	% Including figure files
\usepackage{amsmath}	% Advanced maths commands
\usepackage{amssymb}	% Extra maths symbols

\def\ergs{{\rm erg\,\,s^{-1}}}
\def\cm2{{\rm cm^{-2}}}

\usepackage{color}

\title[Lense-Thirring precession around black holes]{On the different flavours of Lense-Thirring precession around accreting stellar mass black holes } 

\author[S.E. Motta et al. ]{
S.E. Motta$^{1}$\thanks{E-mail: sara.motta@physics.ox.ac.uk},
A. Franchini$^{2}$,
G. Lodato$^{2}$,
G. Mastroserio$^{3}$
\\
$^{1}$University of Oxford, Department of Physics, Astrophysics, Denys Wilkinson Building, Keble Road, Oxford OX1 3RH, UK\\
$^{2}$Dipartimento di Fisica, Universit\`a Degli Studi di Milano, Via Celoria, 16, Milano, I-20133, Italy\\
$^{3}$Anton Pannekoek Institute, University of Amsterdam, Science Park 904, 1098 XH Amsterdam, The Netherlands
}

\date{Accepted XXX. Received YYY; in original form ZZZ}

\pubyear{2015}

\begin{document}
\label{firstpage}
\pagerange{\pageref{firstpage}--\pageref{lastpage}}
\maketitle

\begin{abstract}

Type-C QPOs in X-ray binaries have been often interpreted as a consequence of relativistic Lense-Thirring precession around a spinning black hole and they potentially offer a way to measure black hole spins and masses. 
The connection between relativistic precession and the resulting QPO has been made either in terms of a simplified model involving a single test particle producing the QPO, or in terms of a global model where a geometrically thick accretion flow precesses coherently as a rigid body.
In this paper, we analyse similarities and differences between these two models, sometimes considered as in opposition to each other. We demonstrate that the former is the limiting case of the latter when the radial extent of the precessing flow is very small, and that solid lower limits to the black hole spin can be obtained by considering the test particle model alone. 
We also show that the global precession model naturally accounts for the range of frequencies observed for Type-C QPOs without the need to invoke a truncation of the inner accretion flow before it reaches the innermost stable circular orbit.
Finally, we show that, in order to maintain rigid precession, the thick accretion flow should be radially narrow, and that if it extends beyond $10-10^2$ gravitational radii it aligns with the black hole spin too fast to produce a coherent QPO.
%\SEM{Finally, we studied the conditions for global rigid precession of the thick accretion flow and we found that the maximum radial extent of such disc is determined by its tendency to align with the black hole spin due to the presence of viscosity rather than by the loss of connection between different regions of the flow necessary for global rigid precession to occur.} %Finally, we demonstrate that, as the thick disc becomes wider, it eventually looses connection among its parts and is not able to precess rigidly, possibly resulting in a loss of coherence of the Type-C QPO. %
\end{abstract}


%%%%%%%%%%%%%%%%%%%%%%%%%%%%%%%%%%%%%%%%%%%%%%%%%%

%%%%%%%%%%%%%%%%% BODY OF PAPER %%%%%%%%%%%%%%%%%%

\section{Introduction}

Low-mass black hole X-ray binaries are typically transient systems that alternate between relatively short outbursts and long periods of (X-ray) quiescence. During outbursts most black hole binaries show significant luminosity changes (from $L \sim 10^{30-31}\,\ergs$ in quiescence to $L \sim 10^{38-39}\,\ergs$), and display a ``hysteresis'' behaviour that becomes apparent as q-shaped loops in the so-called Hardness-Intensity Diagram (HID; see e.g., \citealt{Homan2001}). 
Most active black hole Low Mass X-ray Binaries (LMXBs) show four main different states: the Low Hard State (LHS), dominated by non-thermal emission;  the High Soft State (HSS), where the thermal emission from an accretion disc dominates the energy spectrum; the Hard Intermediate State (HIMS) and the Soft Intermediate State (SIMS), where the energy spectrum shows the contributions from both the accretion disc and the non-thermal emission, but where the most dramatic changes in the timing properties are observed. A few source have also shown an {\it anomalous} or Ultra-Luminous State (ULS, \citealt{Motta2012}), which is similar to a SIMS, but characterized by much higher luminosities.
The different states are defined based on the spectral properties of the source and on the inspection of their Fourier power density spectra (PDS), where changes in the fast time variability are most clear. Among the most remarkable features detected in the PDS there are narrow peaks known as quasi-periodic oscillations (QPOs).

Discovered in the '80s in the X-ray signal from accreting stellar mass black hole and weakly magnetized neutron stars, QPOs have been regularly observed in all accreting systems, including \textit{Ultra Luminous X-ray sources} (\citealt{Strohmayer2003}) and a few Active Galactic Nuclei (\citealt{Gierlinski2008}). 
The centroid frequencies of QPOs can be measured with high accuracy and are typically associated with the motion of matter in the accretion flow or to accretion-related time scales. QPOs yield relatively short (sub-second) time scales and simple light crossing arguments indicate that these features must originate in the innermost regions of the accretion flow.

In black hole X-ray binaries, low frequency QPOs (LFQPOs) are very strong and commonly observed features, and  have been divided  into three subclasses: Type-A, B and C (see e.g. \citealt{Wijnands1999}, \citealt{Casella2005},
\citealt{Motta2015}). Type-C QPOs are the most commonly observed, with frequencies spanning the $\sim 0.1$--30 Hz range, tightly correlated with the spectral evolution of the host source (see e.g. \citealt{Belloni2016}). In the LHS, type-C QPOs appear at frequencies from $\lesssim$ 0.1 Hz to $\sim$1 Hz (with frequency increasing with luminosity). As a source evolves through an outburst, moving to softer states, type-C QPOs show increasing frequencies, reaching their maximum ($\sim$ 10-30 Hz, depending on the source) in the HSS. As a source crosses the SIMS, type-C QPOs are substituted by type-B QPOs. During the ULS, type-C QPOs are  often observed again, occasionally simultaneously with a type-B QPO.
After reaching a maximum frequency, type-C QPOs typically disappear (this is what determines the start of the SIMS), to reappear again when the source eventually makes its way back towards the LHS, after crossing again the SIMS and then the HIMS. During this phase, type-C QPOs decrease in frequency down to less than a few Hz and eventually become non-detectable.

\bigskip

The physical origin of QPOs is still debated, despite the fact that they have been studied extensively since their discovery. In particular, several models have been proposed to explain the common type-C QPOs, which are often interpreted as an effect of relativistic (i.e. frame dragging-related) precession within the accretion flow. 
The Relativistic Precession Model is the first model based on relativistic precession and was originally proposed in the late '90s by \cite{Stella1998} and \cite{Stella1999} to explain QPOs in both neutron stars and black hole binaries. This model considers test particle orbits and the associated frequencies calculated in the Kerr metric. In particular,  the type-C QPOs are explained as the effect of the Lense-Thirring precession of a test particle orbiting a spinning black hole. 

\cite{Ingram2009} proposed a model for type-C QPOs based on similar premises. This model requires a cool optically thick, geometrically thin accretion disc  (the \textit{thin disc}, \citealt{Shakura1973}) truncated at a certain radius (\citealt{Esin1997}, \citealt{Poutanen1997}), filled by a hot, geometrically thick accretion flow (the \textit{thick disc}%\footnote{A disc is to be considered geometrically thick when its viscosity parameter $\alpha$ is smaller than the disc aspect ratio $H/R$ \citep{Shakura1973}. }
) that precesses as a consequence of frame-dragging. This geometry is known as the \textit{truncated disc model}. Variations in the truncation radius of the thin disc determine changes in the geometry of the thick disc (i.e. the thin disc inner truncation radius coincides with the thick disc outer radius). Since the Lense-Thirring frequency increases sharply with decreasing radius, in this model the type-C QPO will have increasing frequency for decreasing thin disc truncation radii and \textit{vice versa}. In particular, in the LHS the type-C QPO will be produced by a thick disc with a large outer radius, which will decrease as the source moves through the LHS,  HIMS and SIMS, reaching a minimum in the HSS, thus making the QPO frequency increase in the process.  

In this framework, the spectral evolution of black hole binaries in outburst, the variations in frequency of type-C QPOs, and the clear correlation between these two observational facts are easily explained. The QPO is produced through the modulation of the X-ray flux coming from self-occultation, projected area and relativistic effects (\citealt{Ingram2012}; \citealt{Veledina2013}, \citealt{Ingram2016}) that become stronger with inclination (see \citealt{Motta2015}). This model also naturally explains why a coherent LF QPO can be observed even when the inner flow is thought to be rather extended (tens of $R_{\rm g}$).

%The Lense-Thirring frequency from the precession of the thick disc can be calculated as a global precession frequency (see e.g. \citealt{Liu2002}, \citealt{Fragile2007},  \citealt{Ingram2009}, \citealt{Franchini2016}) by assuming that the thick disc behaves as a rigid body. When the thin disc inner truncation radius tends to the radius of the ISCO, i.e. as the thick disc approaches the shape of a narrow annulus, the global precession frequency produced via rigid Lense-Thirring precession of the inner flow (the \textit{rigid precession frequency}) approaches the Lense-Thirring frequency of a test particle at the ISCO (the \textit{test particle precession frequency}). 

In this paper we present a study of the test particle precession and the global rigid precession. We describe the conditions under which each of these two processes can be used to explain the type-C QPOs, and to infer the properties of the accretion flow from which they arise.

%__________________________________________________________________

\section{Test particle versus rigid precession}

In the Relativistic Precession Model by \citet{Stella1998}, three observed QPO frequencies are associated with the Keplerian frequency, the periastron precession frequency, and the nodal precession frequency related to the Lense-Thirring effect at a given radius in the disc. \cite{Motta2014a} showed that the equations describing these three frequencies, predicted by the theory of General Relativity, form a system of three equations for the three unknowns: $M$ (the black hole mass), $a$ (the dimensionless black hole spin parameter) and $R$ (the radial location at which the QPOs are produced around the black hole). Such system of equations can be solved analytically if three QPOs of the correct type are observed simultaneously, or numerically when an independent measurement of the black hole mass is available (e.g. from dynamical spectro-photometric observations), together with two of the three QPOs relevant for the RPM (see \citealt{Motta2014a}, \citealt{Motta2014b}, \citealt{Ingram2014}). 

\begin{figure}
\centering
\includegraphics[width=0.50\textwidth]{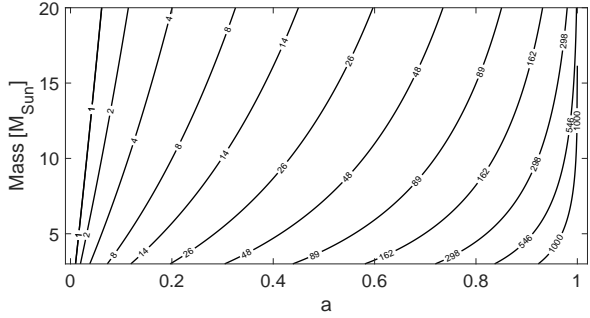}
\caption{Mass-spin relation for different test-particle Lense-Thirring frequencies. Each line has been obtained assuming that the frequency is produced at the ISCO. The number on each line corresponds to the frequency considered in Hz.}
\label{fig:particle}
\end{figure}

In this paper, we are especially concerned with the Lense-Thirring precession frequency (the \textit{test particle precession frequency}) that is given by:
\begin{equation}
\nu_{\mathrm{LT}} = \pm \frac{c^3}{2\pi GM} \frac{1}{r^{3/2} \pm a}\left[ 1 - \left(1 \mp 4ar^{-3/2} +3a^2r^{-2}\right)^{1/2}\right], \label{eq:nod}
\end{equation}
where the top and bottom signs refer to prograde or retrograde precession respectively, and where $r=R/R_{\rm g}$, with $R_{\rm g}=GM/c^2$. Figure \ref{fig:particle} shows the contours of (prograde) $\nu_{\rm LT}$ computed by using Eq. (\ref{eq:nod}) at the ISCO, as a function of mass and spin of the black hole. One can see that $\nu_{\textrm{LT}}$ at the ISCO is of the order of $\approx 1000$ Hz for maximally spinning black holes, with a weak dependence on the black hole mass.

A strong argument against the test particle precession model is that one expects any local disc feature, whose emission is responsible for the QPO, to be rapidly sheared out by differential rotation. The so-called \emph{rigid precession model} (see e.g. \citealt{Ingram2009}) attempts to solve this problem by requiring that type-C QPOs are produced via the \textit{global}, rigid precession of a toroidal structure, rather than being produced via precession of a single test particle. The global \textit{rigid precession frequency} is given by (see e.g. \citealt{Lodato2013,Ingram2012}):
\begin{equation}
\nu_{\mathrm{p}}=\frac{\int_{R_{\rm in}}^{R_{\rm out}}\nu_{\rm LT}(R)L(R)2\pi R\, \mathrm{d}R}{\int_{R_{\rm in}}^{R_{\rm out}}L(R)2\pi R\, \mathrm{d}R},\label{eq:freqP}
\end{equation}
where $R_{\rm in}$ and $R_{\rm out}$ are the inner and outer radii of the rigidly precessing thick disc, $L(R)= \Sigma(R) \Omega_{\rm K}R^2\propto \Sigma(R)R^{1/2}$ is the specific angular momentum,  $\Sigma(R)$ is the radial surface density and $\Omega_{\rm K}$ is given by the relativistic correction to the Keplerian angular velocity.  
Following \cite{Shakura1973}, we will assume a no-torque boundary condition radial surface density of the form: 
\begin{equation}
\Sigma(R)\propto R^{-p}\left(1-\sqrt{\frac{R_{\rm in}}{R}}\right)^p, \label{eq:sigma}
\end{equation}
where the index $p$ varies between 0 and 1. 
The rigid precession frequency is sensitive to the shape of the surface density profile in the vicinity of the ISCO. In particular, non-zero surface densities at the ISCO yield higher rigid precession frequencies, which is a natural consequence of the fact that $\nu_{\rm LT} \propto r^{-3}$. In order to avoid exceedingly high type-C QPO frequencies, \cite{Ingram2009} had to assume a truncated surface density profile. This is not necessary if we a assume a surface density profile in the form given by Eq. \ref{eq:sigma}, which for p = 3/5 corresponds to a radiation pressure dominated Shakura and Sunyaev disc with viscosity proportional to the gas pressure \citep{Franchini2016}.
In this work we will assume that the inner radius $R_{\rm in}$ of the thick disc \textit{always} coincides with the radius of the ISCO (therefore the surface density profile will always be smoothly truncated close to the ISCO), while the outer radius can in principle assume any value larger than $R_{\rm ISCO}$.
Of course, the surface density profile given by Eq. \ref{eq:sigma} does not correspond to the form described in \cite{Shakura1973} if $p\neq 3/5$. However, provided that the profile tends to zero for $R \rightarrow R_{\rm ISCO}$, the rigid precession frequency yielded by Eq. \ref{eq:freqP} does not depend significantly on its specific form, as we will show below.

\begin{figure}
%\centering
\includegraphics[width=0.50\textwidth]{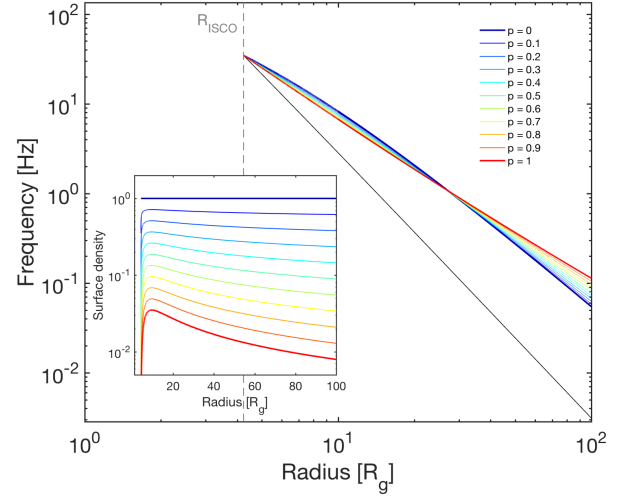}
\caption{Lense-Thirring frequency for a test particle (black solid line) and rigid precession frequency (coloured lines) as a function of radius calculated using a surface density profile in the form: $\Sigma \propto r^{-p} (1 - (R_{\rm in}/r)^{1/2})^p$, with $R_{in} = R_{ISCO}$ (displayed in the inset). In both panels different colours correspond to different values of $p$ (see legend). The frequencies have been calculated for a 10\, M$_{\odot}$ black hole with dimensionless spin parameter $a = 0.5$.}
\label{fig:rigid_vs_tp}
\end{figure}

\begin{figure}
%\centering
\includegraphics[width=0.48\textwidth]{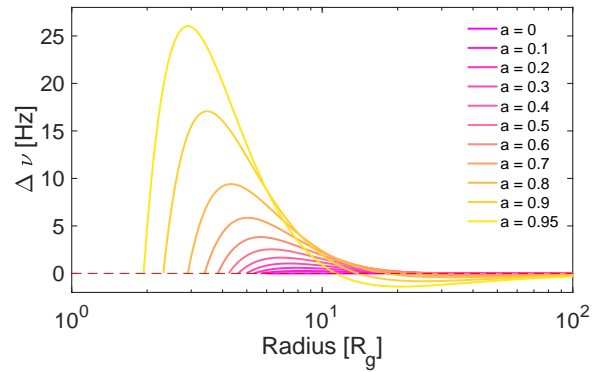}
\caption{Maximum spread as a function of radius for rigid precession frequencies computed for variable spin (see legend) and a 10 M$_{\odot}$ black hole. The maximum frequency spread is calculated as the difference between the rigid precession frequencies computed for $p = 0$ and for $p = 1$.}
\label{fig:deltanu}
\end{figure}

\begin{figure}
%\centering
\includegraphics[width=0.50\textwidth]{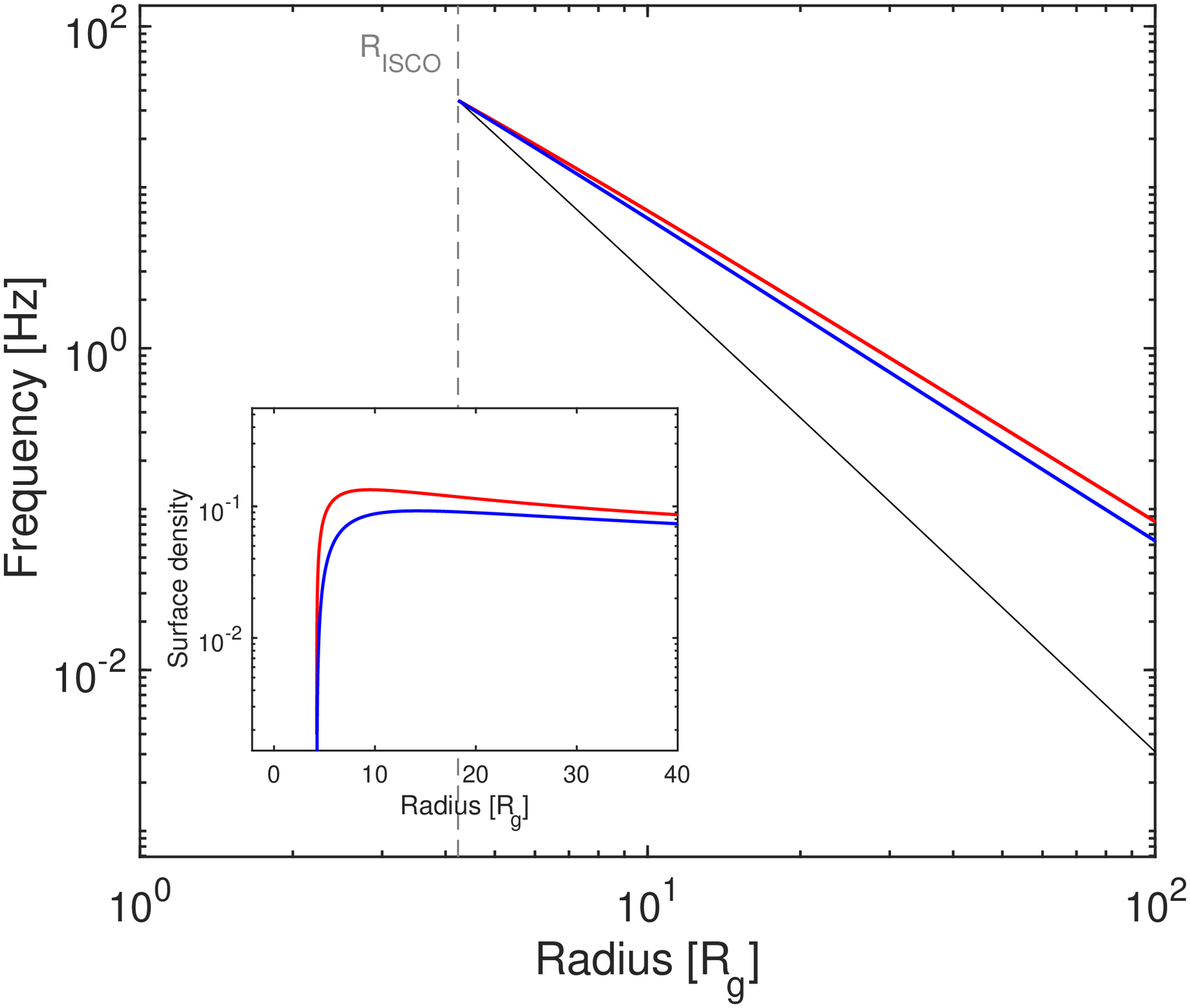}
\caption{Rigid precession frequencies obtained for a $10 M_{\odot}$ black hole, with spin a = 0.5, calculated assuming the surface density profile given in Eq. \ref{eq:sigma} (red line) and those obtained taking a surface density profile with the form $\Sigma(R)\propto R^{-p}\left(1-\sqrt{\frac{R_{\rm in}}{R}}\right)$ (blue line). The inset shows the surface density profiles of the two cases,}
\label{fig:comparison_freq}
\end{figure}

In Fig. \ref{fig:rigid_vs_tp} we show a comparison between the test particle precession frequency given by Eq. (\ref{eq:nod}) and the rigid precession frequencies obtained through Eq. (\ref{eq:freqP}) for a 10 M$_{\odot}$ black hole with spin $a = 0.5$. The rigid precession frequency depends on the inner radius $R_{\rm in}$ (which we fix at $R_{\rm ISCO}$) and on the outer radius $R_{\rm out}$ of the thick disc. The test particle frequency depends only on the radius of the orbit that the test particle is following. For comparison, in the figure this radius is the same $R_{\rm out}$ used to compute the global frequency. The rigid precession frequency depends on the choice of $p$ (indicated with different colours in the plot) through the surface density profile, which is displayed in the figure inset. At large radii, values of $p$ close to zero give a steeper relation between frequency and outer radius than values of $p$ close to one. Note, however, that the dependence on $p$ is generally weak, especially at large radii. The spread in frequency caused by $p$ is spin-dependent and reaches its maximum at radii smaller than $\approx 10 R_{\rm g}$ (note that the plot in Fig. \ref{fig:rigid_vs_tp}  is in log-scale). This effect is better displayed in Fig. \ref{fig:deltanu}, where we plot the difference $\Delta \nu$ between the rigid precession frequencies calculated assuming $p=0$ and $p=1$, as a function of $R_{\mathrm{out}}$ for a 10 M$_{\odot}$ black hole. For spins lower than $\approx$ 0.7 the frequency difference remains smaller than $\approx$ 5 Hz, while it increases quickly for higher spins.

Figure \ref{fig:comparison_freq} shows a comparison between the rigid precession frequencies obtained for a $10 M_{\odot}$ black hole with spin a = 0.5, calculated assuming the surface density profile given in Eq. \ref{eq:sigma}, and the rigid precession frequencies obtained taking a surface density profile with the form $\Sigma(R)\propto R^{-p}\left(1-\sqrt{\frac{R_{\rm in}}{R}}\right)$. In both cases we assumed $p = 3/5$. The maximum difference in frequency using the two different profiles is smaller than 1 Hz at any radius.
As mentioned above, $p = 3/5$ is the most appropriate value for a radiation pressure dominated Shakura-Sunyaev disc, and thus for the thick discs that we consider here, with viscosity proportional to the gas pressure and opacity dominated by electron scattering. Unless otherwise specified, from now on we will therefore assume $p = 3/5$.

\section{Spin measurements from relativistic precession}

As it is clear from Fig. \ref{fig:rigid_vs_tp}, when the outer radius of the thick disc $R_{\rm out}$ approaches the ISCO (i.e. as the thick disc becomes more and more narrow), the rigid precession frequency asymptotically tends to the test particle frequency. 
Based on this property, in \citet{Franchini2017}, we obtained lower limits (and a few upper limits) for the black hole spin in a number of systems by using the highest frequency type-C QPO detected in the soft state in a given source. We assumed that in this state the precessing thick disc is squeezed to a very narrow radial extent, such that we could substitute the (simpler) test particle frequency (Eq. \ref{eq:nod}) to the  rigid precession frequency (Eq. \ref{eq:freqP}) for $R \approx R_{\rm ISCO}$.

We noted, however, that the lower limits to the spins that we obtained were all  significantly below the maximally spinning value $a\sim 1$ (the highest spin we obtained is $a = 0.47$ for the case of 4U 1543-47), despite the fact that the soft-state type-C QPOs that we detected in the HSS spanned a fairly large frequency range ($\sim$ 5-30 Hz). 
The origin of this trend can be deduced from Fig. \ref{fig:particle}, which shows that for black hole masses below $10M_{\odot}$, a spin estimate of $a\gtrsim 0.8$ would correspond to a QPO frequency above 500 Hz, much above the maximum frequency type-C QPOs  observed so far ($\approx 30 Hz$, see e.g.\citealt{Motta2015}). Similarly, Fig. \ref{fig:rigid_vs_tp} shows that for any given radius, the rigid precession frequency is higher than the test particle precession frequency by a factor that we empirically estimate to be $\sim R/R_{\rm ISCO}$. This implies that for a given mass, frequencies explained via simple test particle precession at ISCO will always require spins systematically lower than those explained in terms of global rigid precession. 

The spin distribution of stellar mass black holes is still largely unknown. \citealt{Fragos2015} showed that a constant distribution of black hole spins (obtained through the fitting of the spectral continuum in outbursting black hole low-mass X-ray binaries, see, e.g., \citealt{McClintock2011}) can be obtained by assuming that the natal spin of the black hole in these systems is likely to be initially low, and then increases during the accretion phase. This would result in a roughly constant spin distribution, but requires a strong coupling between the core and the envelope of the progenitor star of the back hole (which has been observed only for red giant stars with mass up to 2M$_{\odot}$). While \cite{Podsiadlowski2003} showed that a significant spin-up during the accretion phase is possible, \cite{King1999} ruled out this possibility  by showing that the amount of mass accreted by the black hole would be too small to significantly change the spin.

Spin measurements obtained via fitting of the reflection features in the energy spectra from accreting black holes seem to indicate that the black hole spin distribution for black hole X-ray binaries might be skewed to high values (see, e.g., \citealt{Reynolds2013}). It must be noted, however, that the spin measurements obtained through this method are typically affected by large uncertainties and, in a few cases, are also in contrast with the measurements obtained through other methods (see, e.g., \citealt{Shafee2006} and \citealt{Reis2009} for the case of GRO J1655-40). This makes it difficult to establish what is the spin distribution underlying the measurements obtained thus far.

Given the above, there is still no obvious reason why certain spin values should be preferred, therefore the simplest assumption we can make is that the spin distribution is somewhat constant. 
In this work we also consider only prograde black hole spins. Even though a few claims of retrograde spin in black hole X-ray binaries can be found in the literature (e.g., \citealt{Reis2013}, \citealt{Morningstar2014}, only a very limited region of the parameter space in a supernova explosion would lead to such a configuration, making a retrograde black hole spin scenario very rare (see, e.g., \citealt{Brandt1995}).
We therefore assumed that the dimensionless spin parameter of a stellar mass black hole can in principle take any value between 0 and 1.
Following this argument, it seems clear that the spin measurements/lower limits based on the test particle precession are likely too conservative, especially if obtained from QPOs that are produced far from the ISCO (\citealt{Franchini2017}).

\begin{figure}
\centering
\includegraphics[width=0.48\textwidth]{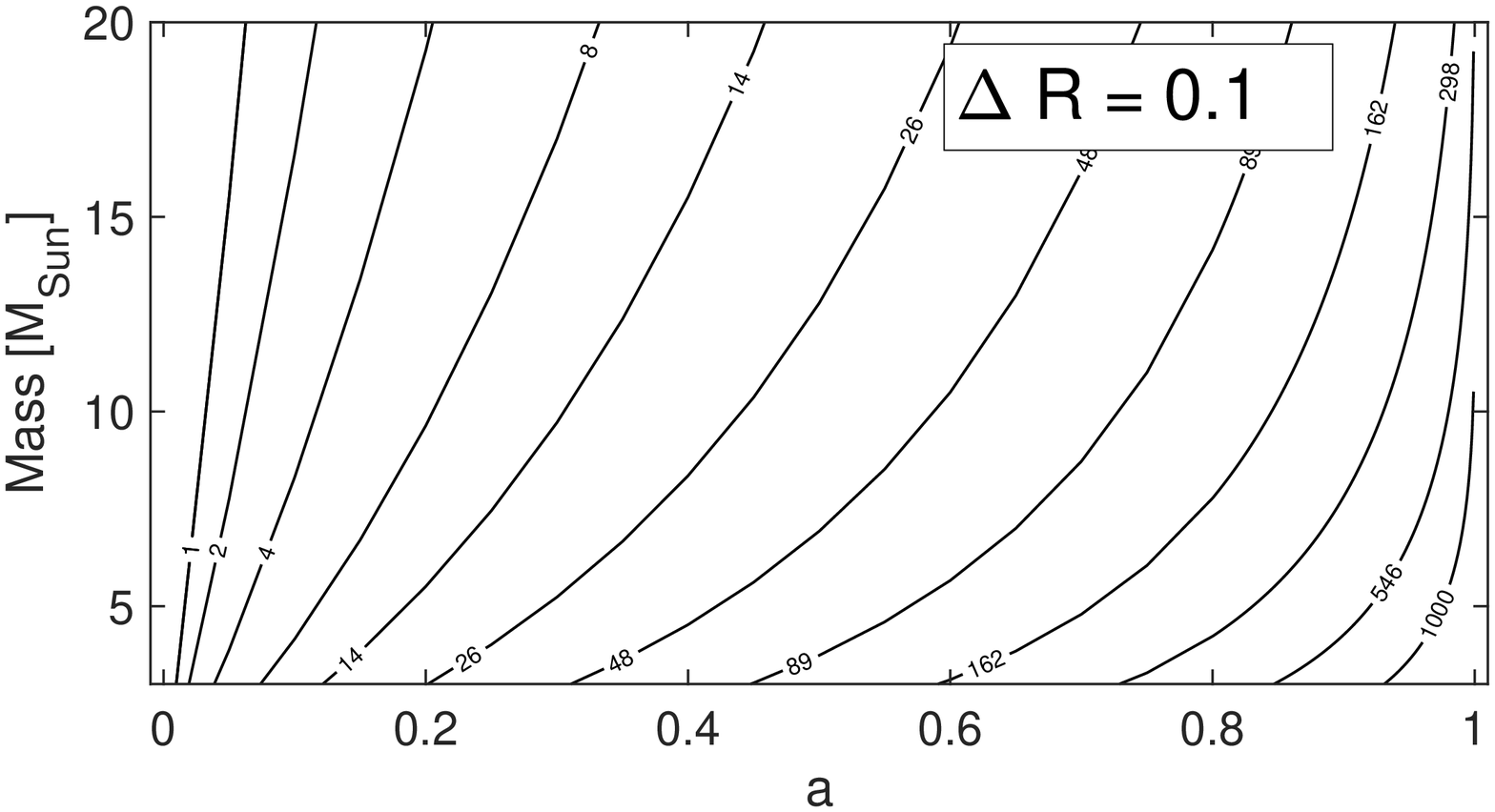}
\includegraphics[width=0.48\textwidth]{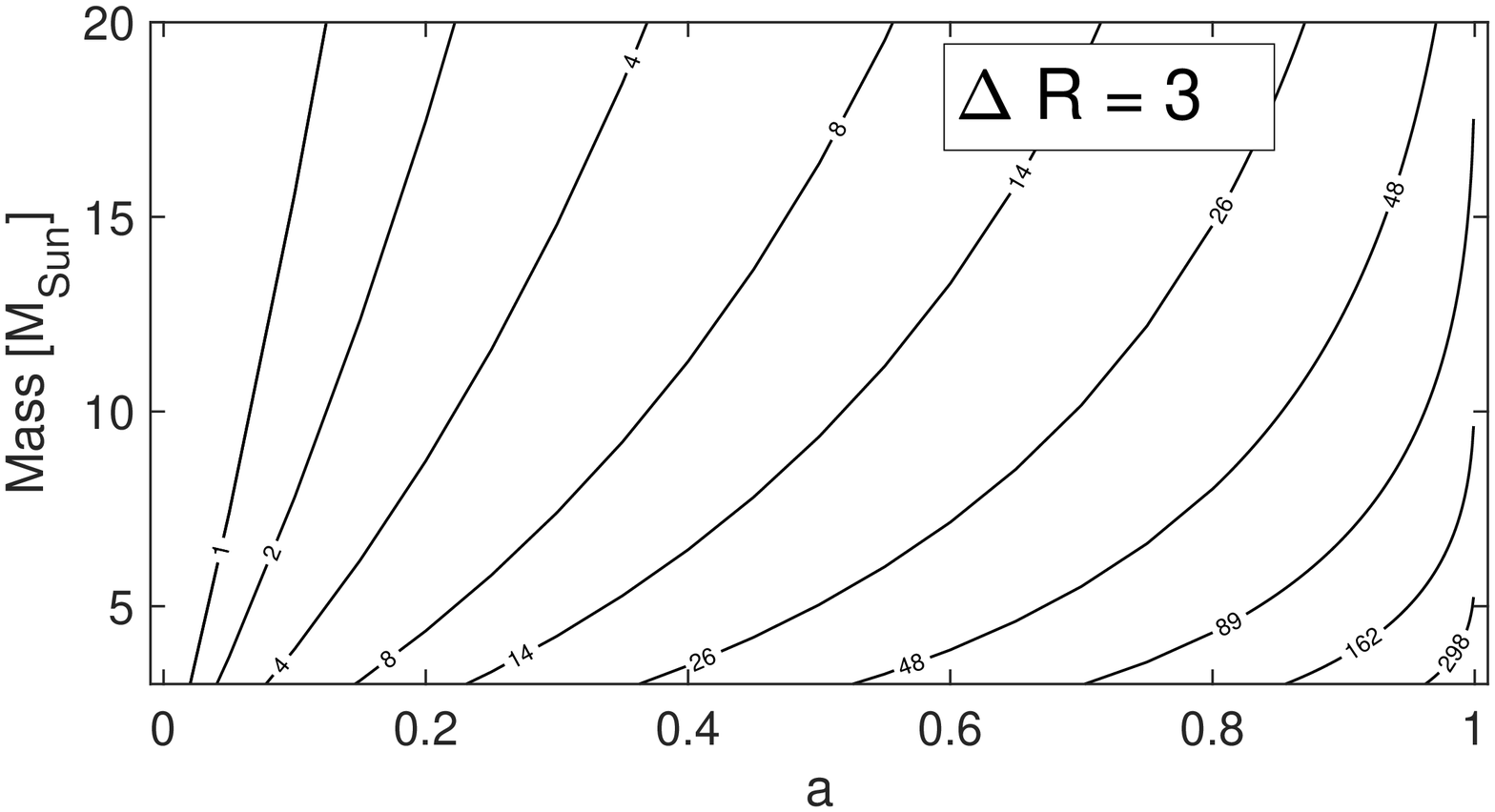}
\includegraphics[width=0.48\textwidth]{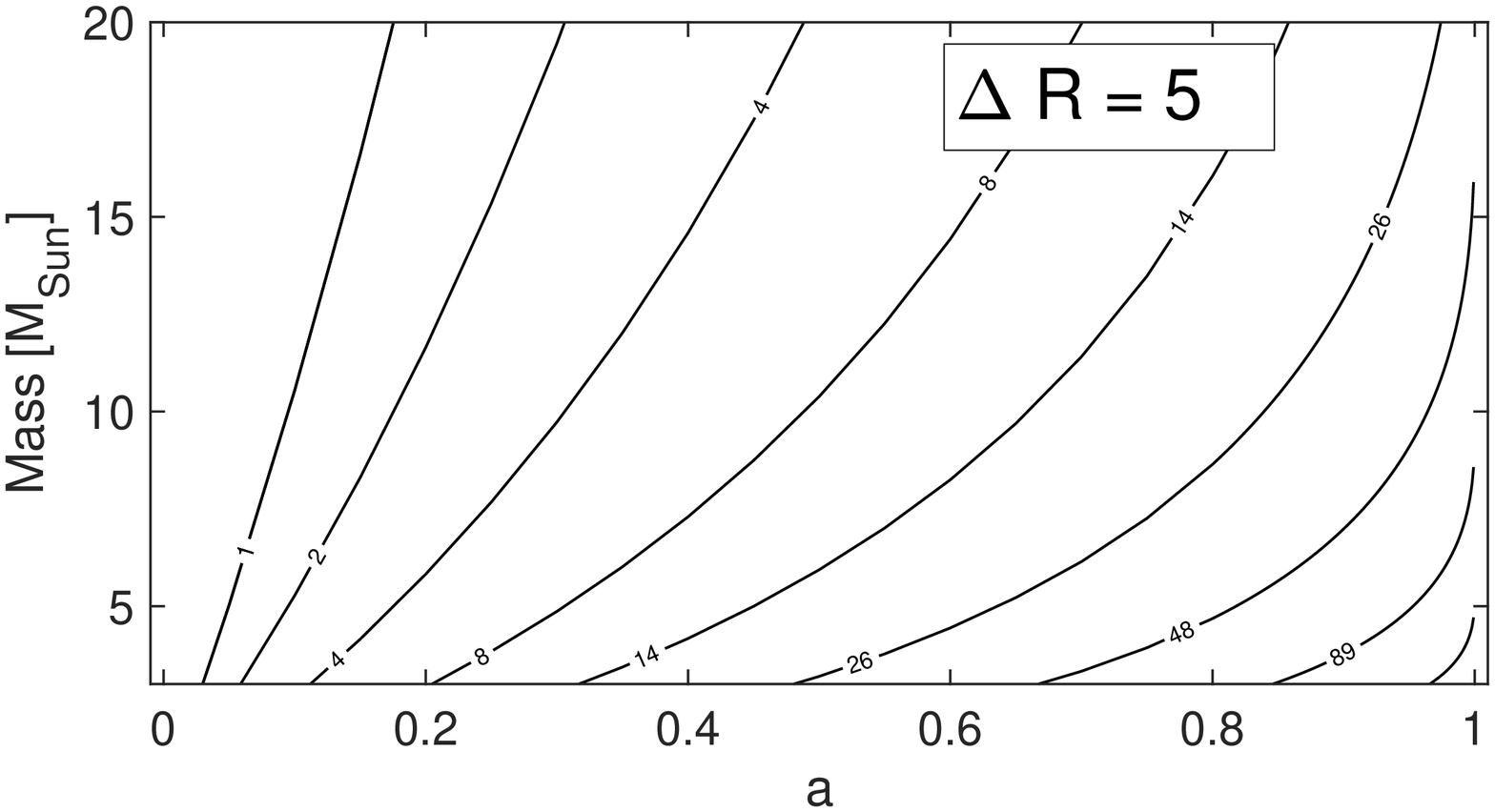}
\includegraphics[width=0.48\textwidth]{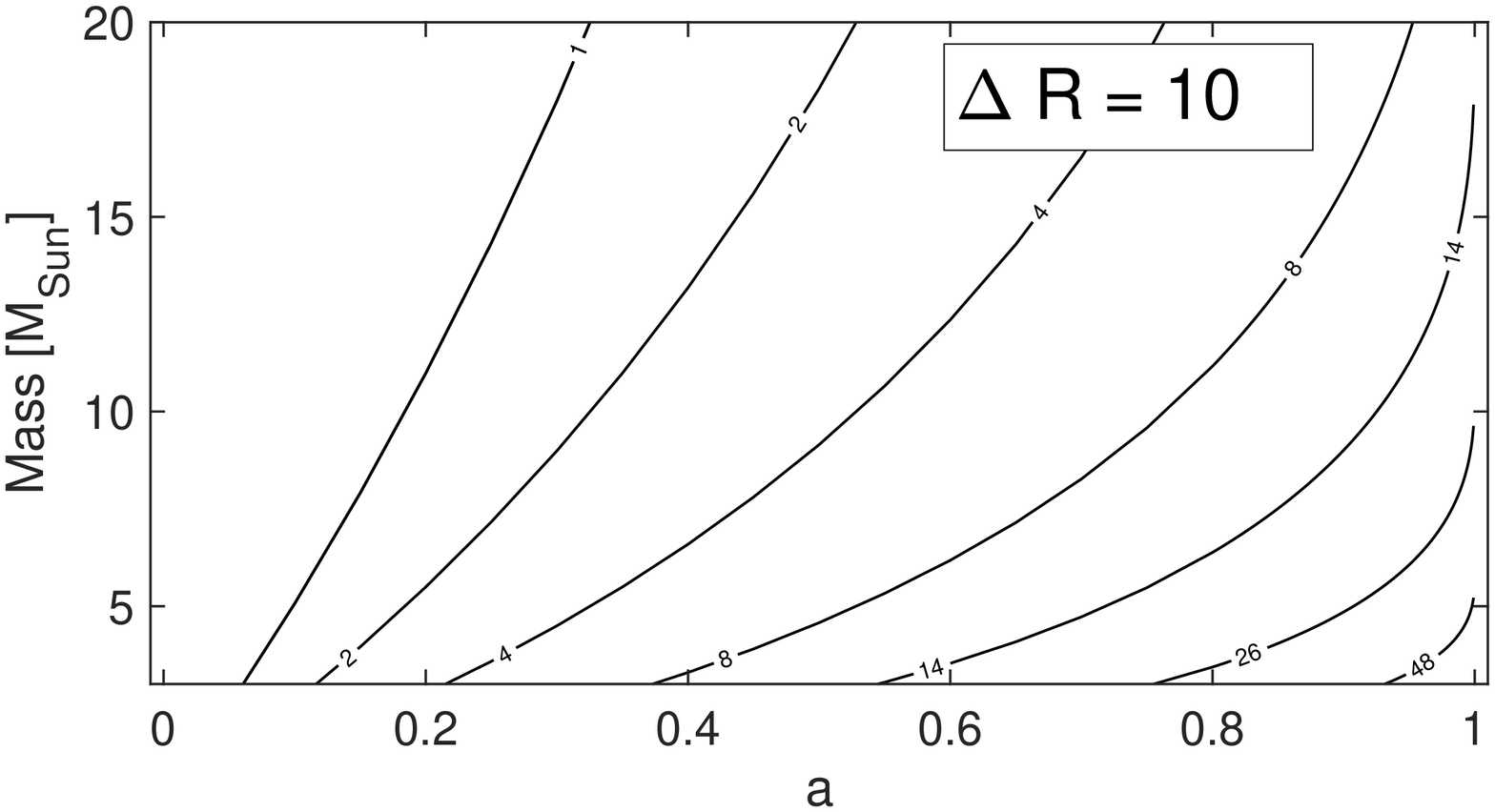}
\caption{Mass-spin relation for different Lense-Thirring frequencies in the rigid precession model. Each line has been obtained assuming that the frequency is produced by a torus extending between $R_{\rm isco}$ and $R_{\rm out}$, with $R_{\rm out} - R_{\rm isco} = \Delta R $ (indicated in each plot). These relations have been obtained assuming the same density profile as in Fig. \ref{fig:deltanu}, with $p = 3/5$.}
\label{fig:mass_vs_radius_rigid}
\end{figure}

An obvious way to limit the maximum global precession frequency is to allow the thick disc to always maintain a finite radial extent. In other words, $R_{\rm out}$ must be always strictly larger the $R_{\rm ISCO}$.  
Figure \ref{fig:mass_vs_radius_rigid} shows contour plots of the rigid precession frequency computed as a function of black hole mass and spin for a number of choices of the outer radius of the thick disc: from top to bottom $\Delta R = R_{\rm out}-R_{\rm in}=$ 0.1, 3, 5 and 10 $R_{\rm g}$. 
While for $\Delta R = 0.1 R_{\rm g}$ the plot is very similar to the test particle case (see Fig. \ref{fig:particle}), for larger thick disc radial extents the resulting frequencies rapidly decrease for a given mass-spin couple. 
We thus see that by allowing a modest radial extent of the thick disc we obtain Lense-Thirring frequencies that are consistent with the type-C QPOs frequencies typically observed around black hole X-ray binaries, with their maximum reaching up to $\approx$ 30 Hz when the thick disc is  a few $R_{\rm g}$ wide (i.e. what we expect in the HSS). Allowing the thick disc to maintain a radial extent implies the entire spin range (0 to 1) must be spanned in order to obtain frequencies typical of type-C QPOs in the HSS for a reasonable range of black hole masses ($\sim 3-20 M_{\odot}$). This essentially removes one of the limitations of the test particle model described above.

\begin{figure*}
\centering
\includegraphics[width=0.32\textwidth]{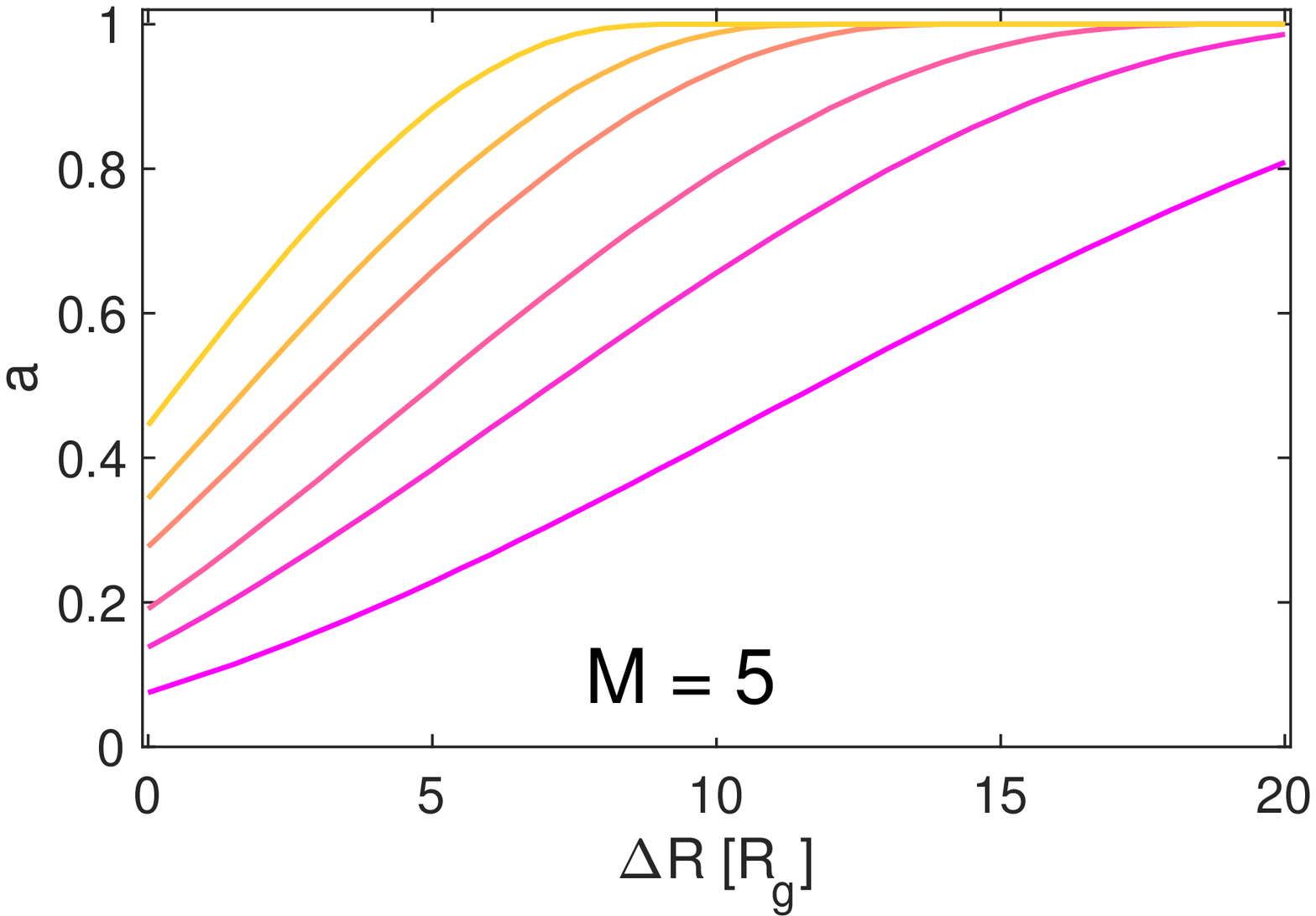}
\includegraphics[width=0.32\textwidth]{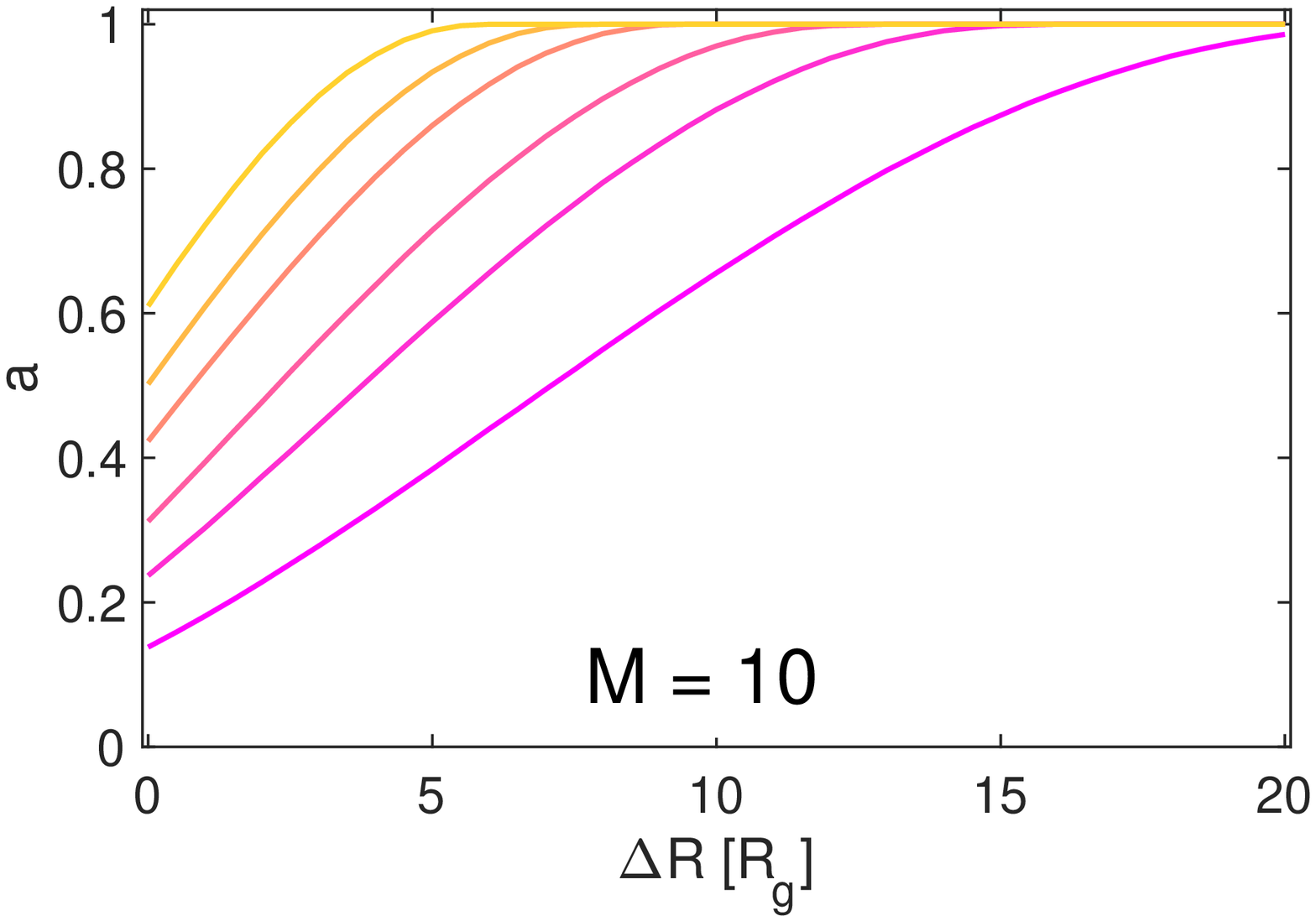}
\includegraphics[width=0.32\textwidth]{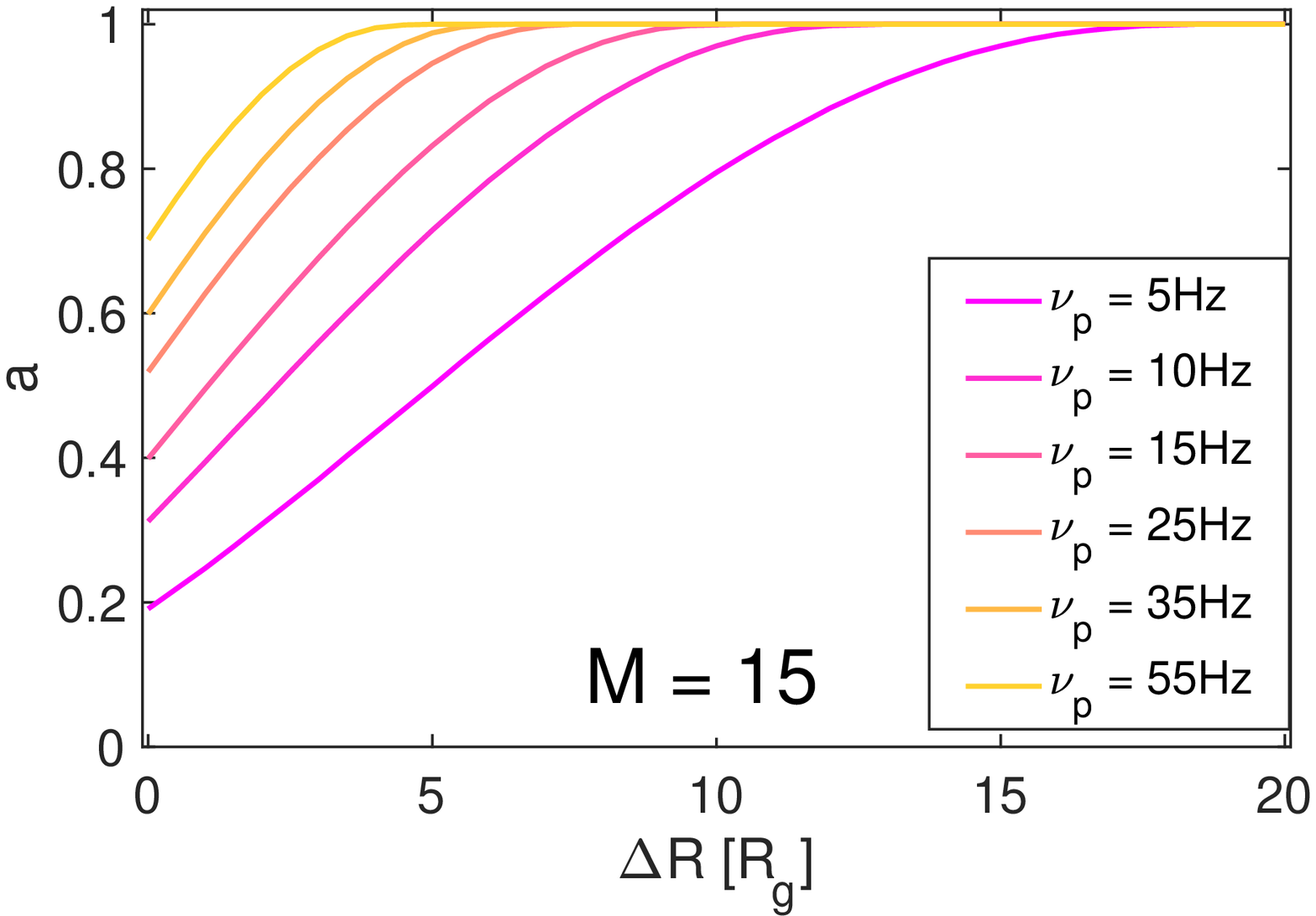}
\caption{Spin inferred for a 5, 10 and 15 M$_{\odot}$ black hole, obtained assuming that a given frequency is produced by a thick disc with increasing $R_{\rm out} - R_{\rm isco} = \Delta R $. In each case, $\Delta R = 0$ corresponds to the test particle precession. The rigid precession frequencies have been computed assuming $p = 3/5$.}
\label{fig:spinvar}
\end{figure*}

Of course, the spin measurement that can be obtained allowing the thick disc to be radially extended depends strongly on its actual radial size. In Figure \ref{fig:spinvar} we plot the value of the black hole spin $a$ obtained from the rigid precession model as a function of the radial extent of the thick disc, assuming a given observed type-C QPO frequency. $\Delta R = 0 $ obviously corresponds to the test particle case. The plots show that the test particle precession provides a solid lower limit to the black hole spin, while a stricter lower limit to the spin (or its actual value) could only be inferred knowing the radial extent of the thick disc reached around a given black hole. 
We note that for wider thick discs the inferred spin quickly ``saturates'' at 1 even assuming  relatively low QPO frequencies, i.e., $\sim$5-15 Hz depending on the mass of the black hole (the higher the mass, the lower the frequency at which the spin saturates at 1). Such regions where the spin is equal to 1 for a variable $\Delta R$ correspond to areas of the parameters space for which there is no solution given a particular frequency and $\Delta R$.
Again under the assumption that the spin distribution of stellar mass black holes is close to a constant in the 0 to 1 range, the above suggests that it is unlikely that the thick disc will have a radial extent $\Delta R$ larger than $\sim$5-10 R$_{\rm g}$ in the HSS.

\begin{figure}
%\centering
\includegraphics[width=0.50\textwidth]{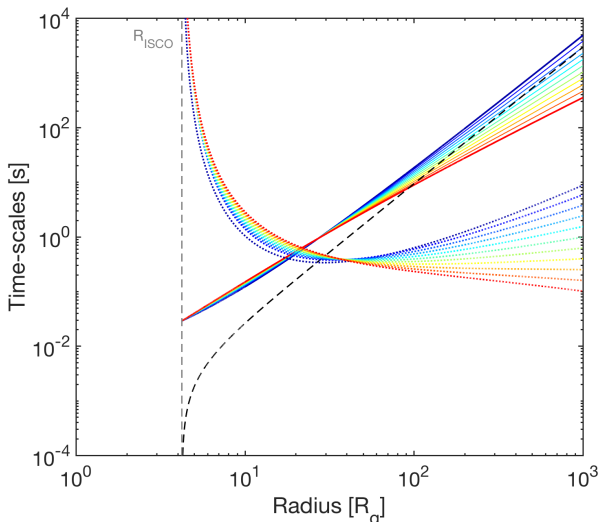}
\caption{Precession time scales (coloured solid lines) compared with the the sound-crossing time scale (black dashed line) and the viscosity-related alignment time scale %(black solid line for $p = 3/5$ and 
(dotted coloured lines)
%for other choices of $p$). 
The solid coloured lines correspond to the precession time scales obtained from the precession frequency shown in Fig. \ref{fig:rigid_vs_tp}. All the time scales have been calculated for a 10\, M$_{\odot}$ black hole with dimensionless spin parameter $a = 0.5$, by assuming $(H/R)_{\rm ISCO}=0.1$ and $q = 3/2$. Each colour corresponds to a different $p$ as in Fig. \ref{fig:rigid_vs_tp}.
}
\label{fig:rigid_condition}
\end{figure}

\section{Requirements for the global rigid precession of the thick disc}\label{sec:rigid}

The outer radius $R_{\rm out}$ of the thick disc %is in principle unconstrained and 
only depends on the physical properties of the accretion flow, and changes depending on the accretion state of a source. 
%However, there exists a maximum radius beyond which the thick disc cannot precess in a rigid manner. In other words, it might happen that only the inner portion of the thick disc can precess rigidly. 
Since, in general, the thick disc angular momentum is not aligned with the black hole spin, the disc undergoes precession due to the Lense-Thirring effect, which induces a precession rate proportional to $R^{-3}$. Thus, the innermost thick disc orbits are subject to a stronger torque and tend to align with the black hole spin while the outer parts maintain the misalignment. This leads to the formation of a warp in the thick disc. In a thick disc the warping disturbances propagate as waves with half the sound speed \citep{Nelson2000}\footnote{In a thin disc the propagation is, instead, diffusive (see, for example, \citealt{LodatoPringle2007}).}. 
Eventually, the thick disc becomes too wide in the radial direction for the warp to be communicated efficiently across it, and at this point rigid precession can no longer occur. Additionally, a very extended disc has the tendency to align with the black hole spin, again inhibiting precession.
Hence, there exists a maximum radius $R_{\rm max}$ beyond which the thick disc cannot precess in a rigid manner. %While $r^{\rm max}_{\rm prec}$ cannot be larger than $r_{\rm out}$, it can be smaller than it, in which case only the inner portion of the thick disc would be able to precess rigidly. 

\subsection{Rigid precession condition}

\cite{Papaloizou1995,Larwood1996,Fragile2005} have shown that the most stringent condition to maintain the rigid precession is that the warp wave must travel
across the thick disc before a precession cycle is over \citep{Larwood1997}. Therefore a simple criterion to ensure that global rigid precession does occur is to require that the precession time scale $t_{\rm p}= 1/\nu_{\rm p}$ is longer than the sound crossing time scale from the inner to the outer radius of the thick disc ($t_{\rm wave}$). This criterion, which we will refer to as \textit{the rigid precession condition}, can be expressed as follows:
\begin{equation}
t_{\rm wave}\lesssim t_{\rm p}, \label{eq:crit}
\end{equation}
with 
\begin{equation}
t_{\rm wave} = \int_{R_{\rm in}}^{R_{\rm out}} \frac{2dR}{c_{\rm s}(R)}.\label{eq:wave}
\end{equation}
The maximum radius $R_{\rm rigid}$ that satisfies the criterion given in Eq. \ref{eq:crit} corresponds to the maximum radial extent that a thick disc can have in order to be able to precess as a rigid body. This, of course, does not impose any limit on the actual size of the thick disc. Rather, it will only determine whether the entire thick disc can precess rigidly.

In order to solve Eq. \ref{eq:crit} we considered a pure power law sound speed profile $c_{\rm s} = c_{\rm s,0}r^{-q}$, with $r = R/R_{\rm g}$. Since $H = c_{\rm s}/\Omega_{\rm K}$, the coefficient of the sound speed evaluated at one gravitational radius is $c_{\rm s,0} = c (H/R)_{0}$, where $c$ is the speed of light.
Solving Eq. \ref{eq:wave}, the wave propagation time scale becomes:

\begin{equation}
t_{\rm wave} = 2\frac{GM}{c^3} \left(\frac{H}{R}\right)_{\rm ISCO}^{-1}r_{\rm ISCO}^{1/2-q}\frac{1}{1+q}\left(r_{\rm out}^{1+q} - r_{\rm in}^{1+q}\right) \label{eq:wavesolved}
\end{equation}  
where $(H/R)_{\rm ISCO} = (H/R)_0 r_{\rm ISCO}^{-q+1/2}$ is the disc aspect ratio at $R_{\rm ISCO}$, and where both $r_{\rm in}$ and $r_{\rm out}$  are expressed in units of $R_{\rm g}$ for convenience. 
Equation \ref{eq:wavesolved} implies that the larger the disc aspect ratio at the ISCO, the shorter the warp propagation time scale. Thus, thicker discs are more likely to precess as rigid bodies than thinner discs.

Figure \ref{fig:rigid_condition} shows the precession time scale and the sound crossing time scale for the case of a black hole with $M=10M_{\odot}$ and $a=0.5$. The coloured solid lines show the rigid precession time scale $t_{\rm p} = 1/\nu_p$ for different values of $p$, the same used in Fig. \ref{fig:rigid_vs_tp}. The maximum outer radius that allows rigid precession, $r_{\rm rigid}$, can be read off as the intersection of these lines with the black dashed line, that marks $t_{\rm wave}$ as defined above (Eq. \ref{eq:wavesolved}), where we have assumed $q=3/2$ and $(H/R)_{\rm ISCO}=0.1$. 
The value of $q=3/2$ is appropriate for radiation pressure dominated discs (\citealt{Frank1992}, Eq. 5.54), such as our thick disc. 
The aspect ratio at ISCO, $(H/R)_{\rm ISCO}$, is %though to be
proportional to $\dot{M}/\dot{M}_{\rm Edd}$ (\citealt{Frank1992}). Since the rigid precession condition is expected to be most relevant where the thick disc is radially very wide  - e.g. in the LHS - we initially chose the value of $(H/R)_{\rm ISCO}$ by considering that in this state, the accretion rate is thought to be low, i.e. $\approx$1-10\%$L_{\rm Edd}$ (\citealt{Fender2014}, \citealt{Dunn2010}), which implies $(H/R)_{\rm ISCO} \sim 0.1$.
If the outer radius $r_{\rm out}$ of the thick disc is larger than $r_{\rm rigid}$, only the section of the thick disc with $r < r_{\rm rigid}$ will be able to precess rigidly. The section of the thick disc with radii larger than $r_{\rm rigid}$ would not able to precess coherently with the inner section of the disc, and could arguably either undergo differential precession (\citealt{vandenEijnden2016}), or break free from the (still precessing) inner flow (see e.g. \citealt{Nealon2015}). 

The rigid precession condition depends on spin of the black hole (while it is independent on the black hole mass $M$, since both $t_{wave}$ and t$_{p}$ are linear functions of $M$), as it directly depends on the precession time scale, as well as on the disc aspect ratio. Since higher spins mean faster precession, higher spins will also correspond to smaller maximum precession radii $r_{\rm rigid}$.
In Fig. \ref{fig:maxradius} we show the behaviour of $r_{\rm rigid}$ (solid lines) as a function of the spin for different values of $(H/R)_{\rm ISCO}$, as obtained through Eq. \ref{eq:wave} for a  spin = 0.5 and mass equal to $10 M_{\odot}$ (for the bottom panel only). %We do not show the dependence of $r^{\rm max}_{\rm prec}$ on the mass of the black hole since the precession time scale is more sensitive to the spin than to the mass of the black hole. 
We see that while for low spins the precession time scale is such that even very wide thick discs will be able to precess\footnote{Note that the entire accretion disc filling the Roche lobe around a stellar mass black hole in a binary system can be as large as 10$^6  R_{\rm g}$, but it is typically smaller than this.} ($r_{\rm out} \gtrsim 10^5 R_{\rm g}$), for very high spins the thick disc will be able to precess globally only as long as its outer radius is lower than $\sim 10 R_{\rm g}$, especially for low aspect ratios. 
We must note, however, that in principle $\dot{M}/\dot{M}_{\rm Edd}$ and, consequently the aspect ratio $(H/R)$, depend on the spin, and both increase for increasing spin. For simplicity we have not considered this effect, however, given the above, the actual maximum precession radii given by Eq. \ref{eq:wave} for a thick discs around a highly spinning black hole are likely larger than those shown in Fig. \ref{fig:maxradius} (solid lines).

\subsection{Condition for disc alignment}\label{sec:align}

While the presence of misalignment between the black hole spin and the disc angular momentum leads to rigid precession, viscosity acts to dissipate the warp, leading to alignment of the disc angular momentum to the black hole spin. If  alignment is achieved faster than one precessional time, rigid precession can no longer occur. 
%The main alignment mechanism is related to the disc 
The disc viscosity damps precession at a rate inversely proportional to the viscosity parameter $\alpha$ ($t_{\rm align}\propto \alpha^{-1}$).
An approximate expression for the damping rate $\gamma \approx t_{\rm align}^{-1}$, based on the estimate of the energy loss rate due to viscous dissipation in the disc, was proposed by \cite{Bate2000} for discs of constant thickness.
\cite{FL2014} give a more accurate estimate of this time scale by computing the difference between the total torque exerted by the black hole on the disc and that required in order to keep the global precession of the disc. The precession damping rate, and therefore the alignment time scale, are given  by: 
%
%\begin{equation}
%
%t_{\rm align} \sim \frac{1}{\alpha} \left(\frac{H}{R}\right)^2 \frac{\Omega}{\Omega_{\mathrm{p}}^{2}} \sim \frac{c^3}{GM} \frac{1}{\alpha} \left(\frac{H}{R}\right)_{\rm ISCO}^2 r_{\rm ISCO}^{2q -1} r^{-2q-1/2}_{\rm out} \frac{1}{\Omega_{\mathrm{p}}^{2}} \,
%\label{eq:tFL}
%\end{equation}
%
\begin{equation}
\gamma = \frac{1}{t_{\mathrm{align}}}= \frac{\int_{R_{\mathrm{in}}}^{R_{\mathrm{out}}} \mbox{d}x\frac{4\alpha {G^2_{\phi}}}{\Sigma {c^2_{\mathrm{s}}} x^3}}{\int_{R_{\mathrm{in}}}^{R_{\mathrm{out}}} \mbox{d}x \Sigma x^3 \Omega_{\rm K}}\label{eq:tFL}
\end{equation}
The term $G_{\phi}$ is the internal stress in the disc and is defined as $G_{\phi}(R) =  \int_{R_{\mathrm{in}}}^{R} \mbox{d}x \Sigma x^3 \Omega_{\rm K} (\Omega_{\mathrm{p}} - Z\Omega_{\rm K})$, where $Z = (\Omega^2_{\rm K} - \Omega^2_{z})/(2\Omega^2_{\rm K})$. 
%
%\begin{equation}
%G_{\phi}(r) =  \int_{r_{\mathrm{in}}}^{r} dx \Sigma x^3 \Omega_{\rm K} (\Omega_{\mathrm{p}} - Z\Omega_{\rm K})\,.\label{eq:gphi}
%\end{equation}
%where $Z$ is defined as $Z(r) = (\Omega^2_{\rm K} - \Omega^2_{\rm z})/(2\Omega^2_{\rm K})$.
% 
In these equations $\Omega_{\rm K}$ is the keplerian orbital velocity, $\Omega_{z}$ the vertical oscillation velocity and $\Omega_{\mathrm{p}} = 2\pi \nu_{\rm p }$ is the global precession angular velocity of the disc. 
%\cite{FL2014} gave a more accurate (but more complex) estimate of this time scale by computing the difference between the total torque exerted by the black hole on the disc and that required in order to keep the global precession of the disc. However, \cite{Franchini2016} showed that such a relation differs from that given in Eq. \ref{eq:tFL} by a factor between 0.25 and 2. Therefore, for simplicity, we will use the condition given above since we are interested in a order-of-magnitude estimate.

%In order to estimate the alignment time scale, we will follow \cite{FL2014}, who found an analytical approximation to the expression given in Eq. \ref{eq:tFL}. According to such approximation, the precession damping frequency - and thus the alignment time scale - is given by:

%\begin{equation}
%\gamma = \frac{1}{t_{\mathrm{align}}} \approx \frac{4 \alpha \Omega_{\rm K,in} Z^2_{\rm in}}{(\frac{H}{R})_{\rm in}^{2}} \Gamma({\rm r_{\rm out}}) \,
%\label{eq:tFL_approx}
%\end{equation}
%
%where $\Gamma$ is a \textit{shape function} that depends on the assumptions made on the disc profiles (see \cite{FL2014}, Appendix A for details) and is a function of $r_{out}$. 

\begin{figure}
\centering
\includegraphics[width=0.48\textwidth]{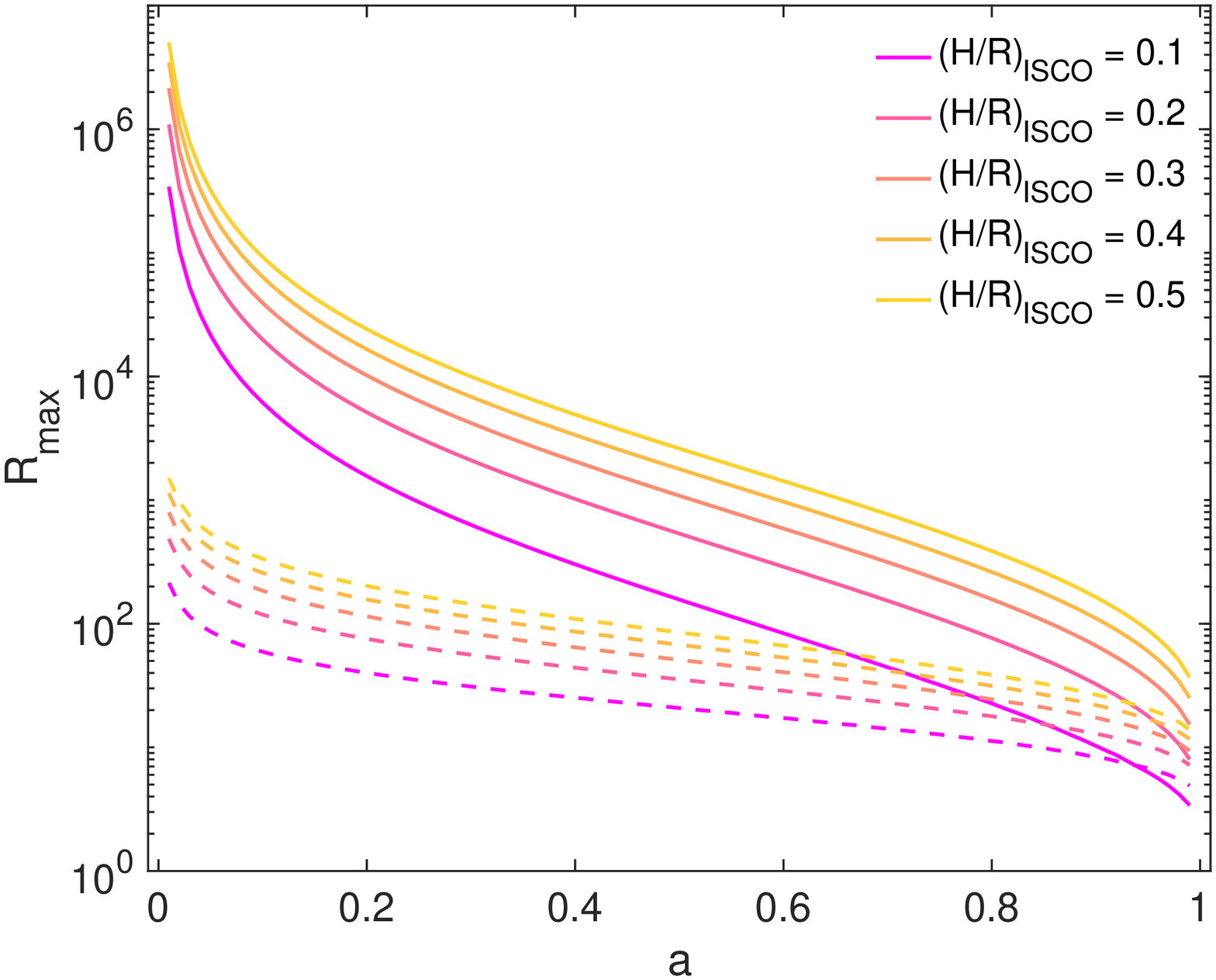}
\includegraphics[width=0.48\textwidth]{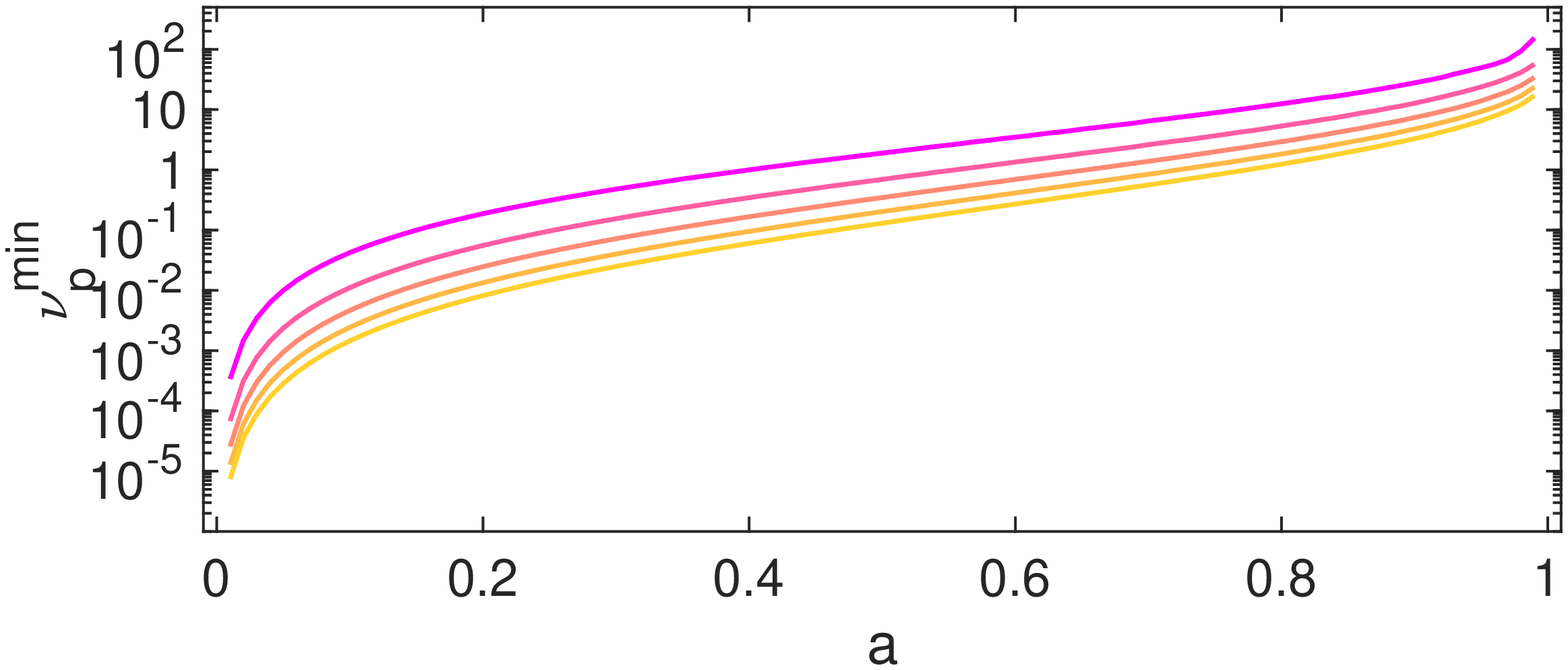}

\caption{\textit{Top panel:} Maximum precession radius $r_{\rm max}$ due to the rigid precession condition (i.e., $r_{\rm rigid}$, solid lines) or the alignment condition (i.e. $r_{\rm align}$, dashed lines) as a function of the spin for variable $(H/R)_{\rm ISCO}$ ($p = 3/5$, $q=3/2$, $\alpha$ = 0.01). The relations displayed have been obtained through Eq. \ref{eq:wave} and \ref{eq:tFL}, respectively. The figure shows that the maximum radius at which global rigid precession is allowed is always determined by the tendency of the disc to align with the black hole spin. 
\textit{Bottom panel:} corresponding minimum global precession frequency as a function of the spins. The colour coding is the same used in the top panel. Frequencies have been estimated assuming a black hole mass of $10 M_{\odot}$.}
\label{fig:maxradius}
\end{figure}

Similarly to the case of the rigid precession condition (Sec. \ref{sec:rigid}), Eq. \ref{eq:tFL} implies that there is a maximum radius $r_{\rm align}$ after which the alignment time scale is faster than the precession time scale. When this happens, the disc aligns to the black hole spin before a complete precession cycle is over, effectively inhibiting continuous precession. Therefore, in order to maintain precession, we also need to require that the thick disc extends no further than $r_{\rm align}$. 
In Figure \ref{fig:rigid_condition} we show the viscosity-related alignment time scales (coloured dotted lines) together with the precession time scales and the sound crossing time scale. Since the alignment time scale ${\rm t}_{\rm align}$ depends on the index $p$ through the surface density profile $\Sigma$, each colour corresponds to a different choice of $p$ (see legend in Fig. \ref{fig:rigid_vs_tp}). %The black solid line marks the alignment time scale calculated for $p=3/5$, that we remind is our default choice for $p$ and will be used for further calculations.
We note that the alignment time scale is shorter for higher black hole spins, % because the ISCO is closer to the black hole and the torque exerted on the disc is thus stronger. This 
which implies that $r_{\rm align}$  becomes smaller for increasing spin. As noted above in the case of the rigid precession condition, we are not considering the effects of the spin on $H/R$, which would likely make $r_{\rm align}$ larger than what obtained simply by solving Eq. \ref{eq:tFL}, especially at high spins. 

It is important to note that in the case of a black hole in a binary system, the binary partner is feeding the compact object, providing a constant stream of misaligned material. This implies that
even though the flow undergoes alignment due to frame dragging within one precession cycle, there is always more matter accreting inclined with respect to the black hole equatorial plane, at least until the system eventually achieves global alignment, which may or may not take longer than the time it takes for the binary partner to evolve.

In a scenario where the alignment time scale is fairly short, precession would still occur, but only intermittently. In other words, precession will only last for the alignment time scale, then will break when the accretion flow reaches alignment, to start again shortly afterwards. In this context, the quality factor Q (defined as $Q = \nu_{\rm QPO} /FWHM_{\rm QPO}$, where $\nu_{\rm QPO}$ is the QPO centroid frequency and $FWHM_{\rm QPO}$ is the QPO full width half maximum) of the resulting QPO would then be approximately the number of precession cycles in an alignment time scale. Since by definition QPOs have Q > 2, this means that we can impose a stricter constraint on $r_{\rm align}$, which is then obtained by requiring $t_{\rm align} > 2 t_p$. 

In Figure \ref{fig:maxradius}, together with the maximum radius at which rigid precession is allowed by the global criterion (Eq. \ref{eq:crit}), we show the maximum radius at which rigid precession is permitted before alignment %in the presence of viscosity 
($r_{\rm align}$, indicated with dashed lines) as a function of the spin. We calculated such radii for variable aspect ratio $(H/R)_{\rm ISCO}$, assuming $\alpha = 0.01$ and $p =3/5$. Interestingly, we see that the disc alignment condition given by Eq. \ref{eq:tFL} almost always dominates over the wave propagation condition given by Eq. \ref{eq:wave}, and therefore the maximum radius $r_{\rm max}=\min(r_{\rm rigid},r_{\rm align})$ at which continuous rigid precession can occur is basically always determined by the tendency of the disc to align with the black hole spin.
%For the sake of completeness, we note that in the context of precession of thick discs around supermassive black holes during tidal disruption events, a second mechanism that could in principle lead to alignment is due to the disc cooling. As a consequence of a decrease in mass accretion rate, the disc cools down and it naturally becomes geometrically thinner, eventually transitioning to a regime where the warp propagates diffusively (\cite{Stone2012}, \cite{Franchini2016}). In this regime the disc should rapidly align with the black hole spin, and the rigid precession should consequently stop accordingly. The alignment time scale related to this mechanism strongly depends on the behaviour of $\dot{M}$ over time, which is largely unknown for accreting black holes in binary systems. Hence, even though this time scale could be in principle relevant in the soft state, where the thick disc could become thinner (see e.g. \citealt{Done2007}), we cannot test how this mechanism would affect rigid precession around a stellar-mass black hole.

\section{Discussion and Conclusions}

We compared the test-particle Lense-Thirring precession and the global rigid precession as likely mechanisms responsible for the presence of Type-C QPOs in the light curve of accreting stellar mass black holes in binary systems.
We showed that the global rigid precession of a radially extended thick disc (as opposed to the test particle precession) can produce QPO frequencies that match those of type-C QPOs typically observed in black hole binaries. This geometry is the same already proposed by other authors, in particular \cite{Ingram2009}, who made the first explicit connection between the idea of rigid precession already observed in numerical simulations (e.g. \citealt{Fragile2007}, \citealt{Liska2017}) and the observed properties of type-C QPOs. 

The geometry that we considered in this work consists of a thick disc whose inner radius is set by the ISCO. In agreement with the truncated disk model, the outer radius of the thick disc, instead, is set by the inner truncation radius of the thin disc concentric and surrounding it, and is always significantly larger than $R_{\rm ISCO}$ (i.e. the radial extent of the thick disc must never be negligible). 
%\GL{Qua ho tolto un pezzo che secondo me confonde le idee inutilmente.}
%Such outer radius never explicitly enters our calculations, which instead involve a  maximum radius (larger or equal to the outer radius of the thick disc) determined by certain rigid precession conditions (see below). 
The surface density profile we assumed is a power-law with index $p$, with a correction at small radii, to satisfy the no-torque condition $\Sigma=0$ at $R=R_{\rm in}$ \citep{Shakura1973}. The index $p$ can be determined based on disc energetics arguments. In the case considered here, we assume that the disc is radiation pressure dominated with viscosity proportional to gas pressure, in which case $p=3/5$.

We showed that in order to match the observed type-C QPOs frequencies it is not necessary to assume that the thick accretion disc is truncated on the inside at a radius larger than the ISCO, contrary to what has been proposed by other authors (e.g., \citealt{Ingram2009}). 
%In general, for a given source, for which the values of mass and spin of the black hole are fixed, the rigid precession frequency is expected to increase (decrease) \emph{in a predictable way} as the inner precessing flow becomes narrower (wider) (see Fig. \ref{fig:rigid_vs_tp}), that is, as the source transitions from the hard to the soft state (from the soft to the hard state). Given a correlation between thin disc truncation and either time or luminosity, this would provide an immediate test for the rigidly precessing model.
Instead, we assumed that while in the HSS the thick accretion disc is  significantly radially narrower than in the other states, its radial dimension should never become negligible. 
We showed that in order to both match the observed type-C QPOs frequencies and span the entire spin and mass range suitable for accreting stellar mass black holes the thick accretion disc must always maintain a non-negligible, though moderate, radial extent, i.e. between a few and $\approx 10 R_{\rm g}$. Additionally, we confirmed that the test particle Lense-Thirring precession applied to the type-C QPOs observed in the HSS allows us to infer a solid lower limit to the black hole spin (see also \citealt{Franchini2017}).

The prediction that the thick disc should maintain a significant radial width is consistent with a well-known property of the PDS of accreting black hole binaries. PDS showing type-C QPOs can be modelled by multiple Lorentzians (see, e.g., \citealt{Belloni2002}). The lowest frequency Lorentzian is generally zero-centered and shows a break at a certain frequency $\nu_{\rm b}$, corresponding to the so-called \textit{low-frequency break} in the flat-top noise typically associated with type-C QPOs. 
The break frequency $\nu_{\rm b}$ is thought to be associated with the viscous time scale $t_{\rm visc} = (H/R)^{-2}(\alpha \Omega)^{-1}$ at the outer radius of the thick disc (see, e.g., \citealt{Done2007}, \citealt{Ingram2011}, ). This frequency is approximately $\sim$ 0.01 Hz in the hard state and $\sim$ 1 Hz in the soft state (see, e.g., \citealt{Done2007}) and roughly correspond to a few tens of $R_{\rm g}$ and a few $R_{\rm g}$, respectively, for a $10 M_{\odot}$  black hole with spin $a=0.5$ (assuming $\alpha$ = 0.01 in the disc), in agreement with our predictions on the disc radial extent in the soft state.

\bigskip
%\AF{The two requirements that need to be satisfied in order for the thick disc to rigidly precess and produce detectable QPOs are the efficiency of the warp propagation and reflection that allow global precession, together with an alignment time scale long enough.} \AF{The rigid precession criterion depends on  the sound crossing time while the alignment time is related to the presence of viscosity in the disc, which leads to the alignment of the disc total angular momentum with the spin of the black hole.}

We then studied the conditions under which the thick disc can undergo global rigid precession around a spinning black hole. The two requirements that need to be satisfied in order for the thick disc to rigidly precess - and thus produce detectable QPOs - are that the disc warp propagates and reflects efficiently, and that the disc alignment time scale is longer than the precession time scale. The first requirement depends on the sound crossing time scale within the disc and guarantees that the warp is maintained. The second requirement depends on the presence of viscosity and ensures that each precession cycle is over before the disc aligns with the black hole spin.
Both conditions depend on the spin of the black hole, and on the disc aspect ratio. The condition based on the alignment time scale depends also on the viscosity parameter $\alpha$. We have hence calculated the maximum outer radius that a thick disc can have in order to both maintain global rigid precession and to keep the misalignment with respect to the black hole spin. 
Interestingly, comparing the maximum precession radii coming from the above conditions% ($r^{\rm max}_{\rm rigid}$ and $r^{\rm max}_{\rm align}$)
, we found that continuous global rigid precession is always inhibited by the disc tendency to align with the spin rather than by the loss of connection between the different regions of the thick disc. In other words, beyond a certain radius $r_{\rm max}$ the disc aligns faster than it precesses, which effectively inhibits precession. Since the global precession frequency decreases with increasing outer radius, this implies that, if type-C QPO are indeed due to global precession, for any given black hole mass there is a minimum allowed QPO frequency, that for a $10M_{\odot}$ black hole is in the range of $0 - 10^{2} {\rm Hz}$ (see Fig. \ref{fig:maxradius}, bottom panel).
Since the vast majority of the known black holes low-mass X-ray binaries show QPOs below 1 Hz, this result essentially implies that none of such black holes can have a spin larger than $a \approx$ 0.6. %These are approximate values obtained in a generic case, therefore we caution the reader not to take them literally, but as indications.

We remind the reader that both the maximum precession radii given by the rigid precession condition and the maximum precession radii given by the presence of alignment are likely higher than the values obtained from Eq. \ref{eq:wave} and Eq. \ref{eq:tFL} for high spin values. As a consequence, the minimum allowed precession frequency is lower than those shown in Fig. \ref{fig:maxradius} (bottom panel), especially for high spins. Furthermore, we note that very recently \cite{Liska2017} have found evidence in a General Relativistic 3D MHD simulation that the jet can play a role in aligning the accretion flow. In particular, the presence of a jet might cause both the precession time scale and the alignment time scale to slow down. Studying further the effects of high spins on the disc dynamics is beyond the scope of this work, and further investigations are left to a future work.

The existence of a maximum radius at which global rigid precession can occur has consequences for the production of type-C QPOs in different accretion states. While in the HSS (and in the ULS) the radial extent to the thick disc is thought to be small, in the LHS the thick disc could be in principle several tens of $R_{\rm g}$ wide. As a consequence, the rigid precession conditions described above could significantly influence the properties of QPOs observed in this state. For instance, a QPO could show a loss in coherence if produced in a thick disc with  $r_{\rm out} \sim r_{\rm max}$ with respect to those QPOs produced when $r_{\rm out} < r_{\rm max}$. This might explain why type-C QPOs detected in active sources caught shortly after they left the quiescence and/or close to their return to it at the end of an outburst. Such QPOs typically appear at frequencies lower than $\sim$ 0.1 Hz and are in general broader and less coherent ($Q \approx 2$) than those observed at higher frequencies (typically $Q \gg 2$).

%In the HSS, instead, the rigid precession condition should not matter much as the thin disc truncation radius is believed to be small, which implies that the thick disc is likely radially narrow as we discussed above. Interestingly, type-C QPOs observed in the HSS or during the ULS (where the thick disc should be significantly more radially compact than in the LHS), they show high coherence ($\nu /\Delta \nu >> 2$, often exceeding 10, \citealt{Motta2012}). 
%We note that in the HSS the accretion rate at ISCO (and so the aspect ratio $(H/R)_{\rm ISCO}$) is expected to be significantly larger than in the LHS ($\dot{M}/\dot{M}_{\rm Edd} \approx 0.5 \approx (H/R)_{\rm ISCO}$ or higher, \citealt{Munoz-Darias2014}). Therefore the maximum precession radius  $r^{\rm max}_{\rm prec}$ is bound to be  larger than in the case of $(H/R)_{\rm ISCO} \approx 0.1$ described in Sec. \ref{sec:rigid}.
	
\bigskip

%\SEM{
%$\Sigma \propto \nu^{-1} \dot{M} [ 1 - \sqrt(Rin/R) ]$,
%and using $\alpha$ prescription, this turns into
%$\Sigma \propto \alpha^{-1} (H/R)^{-2} R^{-0.5} \dot{M} [ 1 - \sqrt(Rin/R) ].$
%}

\bigskip

\section*{Acknowledgements}

We would like to thank the anonymous referee, whose comments and suggestions helped improving this paper. We also thank Chris Fragile and the members of the \textit{``There it spins: the hunt for black hole spins''} ISSI team (call 2014), for useful discussion. We acknowledge ISSI Bern for hospitality and financial support. SEM acknowledges the Violette and Samuel Glasstone Research Fellowship program and the STFC for financial support.

%-------------------------------------------------------------------

\bibliographystyle{mnras.bst}
\bibliography{biblio.bib} 

\end{document}